\documentclass[aps,prb,showpacs,longbibliography,superscriptaddress,citeonline,twocolumn]{revtex4-2}
\usepackage{amsmath,amssymb,bm}
\usepackage{graphicx}
\usepackage{xspace}
\usepackage{latexsym}
\usepackage{xcolor}
\definecolor{LinkColor}{rgb}{0.256,0.439,0.588}
\usepackage{hyperref}
\hypersetup{
colorlinks=true,
citecolor=LinkColor,
linkcolor=LinkColor,
urlcolor=LinkColor
}
\usepackage{orcidlink}

\begin{document}

\title{Critical Berezinskii-Kosterlitz-Thouless dynamics in the archetypal two-dimensional spin system Ba$_2$CuSi$_2$O$_6$Cl$_2$}
\author{K.~M.~Ranjith\,\orcidlink{0000-0001-8681-2461}}
\email{ranjith.km1857@gmail.com}
\affiliation{
Laboratoire National des Champs Magn\'{e}tiques Intenses, LNCMI-CNRS (UPR3228), EMFL, \\
Univ. Grenoble Alpes, Univ. Toulouse, INSA-T, 38042 Grenoble Cedex 9, France}
\affiliation{Present address: Center for Quantum Technologies and Complex Systems (CQTCS),
Christ University, Bengaluru, Karnataka 560029, India}
\author{Maxime Dupont\,\orcidlink{0000-0001-5719-5687}}
\affiliation{Univ. Toulouse, CNRS, Laboratoire de Physique Th\'eorique, Toulouse,  France.}
\affiliation{Present address: Rigetti Computing, 775 Heinz Avenue, Berkeley, California 94710, USA}
\author{Steffen Kr\"{a}mer\,\orcidlink{0000-0002-6107-3583}}
\affiliation{
Laboratoire National des Champs Magn\'{e}tiques Intenses, LNCMI-CNRS (UPR3228), EMFL, \\
Univ. Grenoble Alpes, Univ. Toulouse, INSA-T, 38042 Grenoble Cedex 9, France}
\author{Sylvain Capponi\,\orcidlink{0000-0001-9172-049X}}
\affiliation{Univ. Toulouse, CNRS, Laboratoire de Physique Th\'eorique, Toulouse,  France.}
\author{Edmond Orignac\orcidlink{0000-0002-3405-9508}}
\affiliation{Univ. Lyon, ENS de Lyon, CNRS, Laboratoire de Physique, F-69342 Lyon, France}
\author{Nicolas Laflorencie\,\orcidlink{0000-0001-8634-1006}}
\affiliation{Univ. Toulouse, CNRS, Laboratoire de Physique Th\'eorique, Toulouse,  France.}
\author{Nobuyuki Kurita\,\orcidlink{0000-0002-7857-8043}}
\affiliation{Department of Physics, Institute of Science Tokyo, Meguro, Tokyo 152-8551, Japan}
\author{Hidekazu Tanaka\,\orcidlink{0000-0001-8235-8783}}
\affiliation{Department of Physics, Institute of Science Tokyo, Meguro, Tokyo 152-8551, Japan}
\affiliation{Present address: Center for Entrepreneurship Education, Institute of Science Tokyo, Yokohama 226-8502, Japan}
\author{Mladen Horvati{\'c}\,\orcidlink{0000-0001-7161-0488}}
\email{mladen.horvatic@lncmi.cnrs.fr}
\affiliation{
Laboratoire National des Champs Magn\'{e}tiques Intenses, LNCMI-CNRS (UPR3228), EMFL, \\
Univ. Grenoble Alpes, Univ. Toulouse, INSA-T, 38042 Grenoble Cedex 9, France}
\date{\today}

\begin{abstract}
We study the spin dynamics in the quasi-2D spin-$1/2$ dimer compound Ba$_2$CuSi$_2$O$_6$Cl$_2$, which exhibits a magnetic field-induced Bose-Einstein condensate (BEC) of triplons. Using nuclear magnetic resonance spin-lattice relaxation rate ($T_1^{-1}$) measurements combined with large-scale quantum Monte Carlo (QMC) simulations, we investigate critical fluctuations across the field-temperature phase diagram. Bridging the behavior observed in 1D and 3D systems, the $T_1^{-1}$ relaxation rate shows a pronounced peak extending well above the N\'eel temperature $T_N$, indicating strong two-dimensional Berezinskii-Kosterlitz-Thouless (BKT)-type fluctuations. A quantitative match between experimental and theoretical BEC phase boundaries validates an effective XXZ model. The study determines the intrinsic BKT transition temperature $T_{\mathrm{BKT}}$ from QMC, revealing a nearly field-independent $T_{\mathrm{BKT}}/T_N \approx 0.74$. Scaling analysis of the relaxation rate shows critical exponents consistent with 2D universality, and a narrow temperature window is identified where 2D physics dominates. These findings establish Ba$_2$CuSi$_2$O$_6$Cl$_2$ as a model system for exploring BKT dynamics in quantum magnets.
\end{abstract}
\maketitle

\section{Introduction}
One of the most remarkable properties of low-dimensional systems is that {\color{black} when they have a  gapless regime, this entire regime is characterized by critical behavior.} Due to the unavoidable finite three-dimensional (3D) coupling of their low-dimensional elements, the corresponding quasi-1D and quasi-2D antiferromagnetic (AFM) quantum spin materials exhibit a 3D-ordered low-temperature phase between the two critical magnetic fields $H_{c1}$ and $H_{c2}$, which can be described in the framework of Bose-Einstein condensation (BEC) \cite{Zapf2014}. These compounds, and in particular their low-energy spin-dynamics  measured by the nuclear magnetic resonance (NMR) spin-lattice relaxation rate $T_1^{-1}$ have been instrumental in defining the appropriate theoretical description. While the phase boundary of their BEC phases is, of course, characterized by the 3D-critical fluctuations, these fluctuations are limited to the immediate vicinity, within 10--20\%, of the critical (N\'{e}el) temperature $T_N$, and at higher temperatures we observe low-D spin dynamics.

In 1D, this {\color{black}regime} has been fully quantitatively accounted for based on the purely 1D Tomonaga-Luttinger liquid (TLL) description \cite{TheBook}, corrected by the random phase approximation (RPA) to take into account the effect of 3D couplings \cite{Klanjsek2008,Bouillot2011,Dupont2018,Horvatic2020}. The purely 1D-critical fluctuations, whose ``origin'' is the zero-temperature point corresponding to the absence of order at any finite temperature, provide the critical power-law dependence that extends very far in temperature ($T$). In the \mbox{1.2$T_N$--3$T_N$} range, there is a visible enhancement of spin fluctuation due to the proximity of the phase transition, which is described by the RPA corrections: $T_1^{-1}(T) \propto T^{1/2K-1}\Phi(K,T_N/T)$, where $K$ is the TLL interaction parameter and $\Phi$ is the enhancement function defined in Refs.~\cite{Dupont2018,Horvatic2020}. The enhancement is modest, by a factor of $\sim$2, leading to relatively small $T_1^{-1}(T)$ peak at $T_N$  \cite{Horvatic2020}.

In quasi-2D compounds, {\color{black} we focus here on the non-frustrated spin systems that are representative for the Berezinskii-Kosterlitz-Thouless (BKT) transition and related spin dynamics. The peak of $T_1^{-1}$ observed in these systems at $T_N$ is much higher than in quasi-1D compounds, the peak enhancement being on the order of 10. The peak} extends at least up to $2T_N$ \cite{Opherden2023}, which is much more than in 3D compounds but less than the extension of the power-law regime in quasi-1D compounds. Above $\approx$1.2$T_N$, we expect that the $T_1^{-1}(T)$ dependence reflects the 2D critical fluctuations and, indeed, the available experimental data have been successfully fitted by a corresponding formula based on the scaling properties \cite{Borsa1992,Opherden2023}. However, this formula depends on the BKT ordering temperature $T_{\textrm{BKT}}$ of the corresponding purely 2D system, a parameter which is \emph{a priori} not known and is very difficult to estimate, and the domain of validity of the formula has not been confronted with any theoretical predictions yet. To address these issues and learn more about spin dynamics in quasi-2D compounds, here we present extensive $T_1^{-1}$ data obtained in an archetypal 2D $S = 1/2$ spin-dimer compound Ba$_2$CuSi$_2$O$_6$Cl$_2$, confronted with the corresponding, state-of-the-art quantum Monte Carlo (QMC) simulations.

\section{Material, model, phase diagram}

Loosely speaking, the parent compound of Ba$_2$CuSi$_2$O$_6$Cl$_2$ studied here is the famous Han-purple compound, BaCuSi$_2$O$_6$, originally hoped to be an ideal and the most simple spin{\color{black}-dimer} compound to represent 2D physics: built from planes hosting tetragonal array of Cu spin-1/2 dimers, where the interplane coupling was supposed to be strongly reduced by the geometrical frustration of the AFM exchange couplings \cite{Jaime2004,Sebastian2006}. However, it turned out that the low-temperature structure of this compound is much more complicated \cite{Allenspach2020}, having as the main shortcoming the stacking of three different types of planes, which altogether leads to physics that cannot be considered simple and/or archetypal. This was successfully removed by doping: In the Ba$_{0.9}$Sr$_{0.1}$CuSi$_2$O$_6$ compound, substituting 10\% of Ba by Sr reinforces an average structure where all the planes are equivalent \cite{Allenspach2021}. However, it is hard to believe that the spin dynamics of this compound is unaffected by its disordered structure. Furthermore, both Han purple and its Sr-doped version have very high critical fields, $H_{c1} \approx$ 23~T and $H_{c2}$ is above the currently available maximum dc magnetic field of 45~T.

\begin{figure}[t!]
	\centering
	\includegraphics[width=\columnwidth]{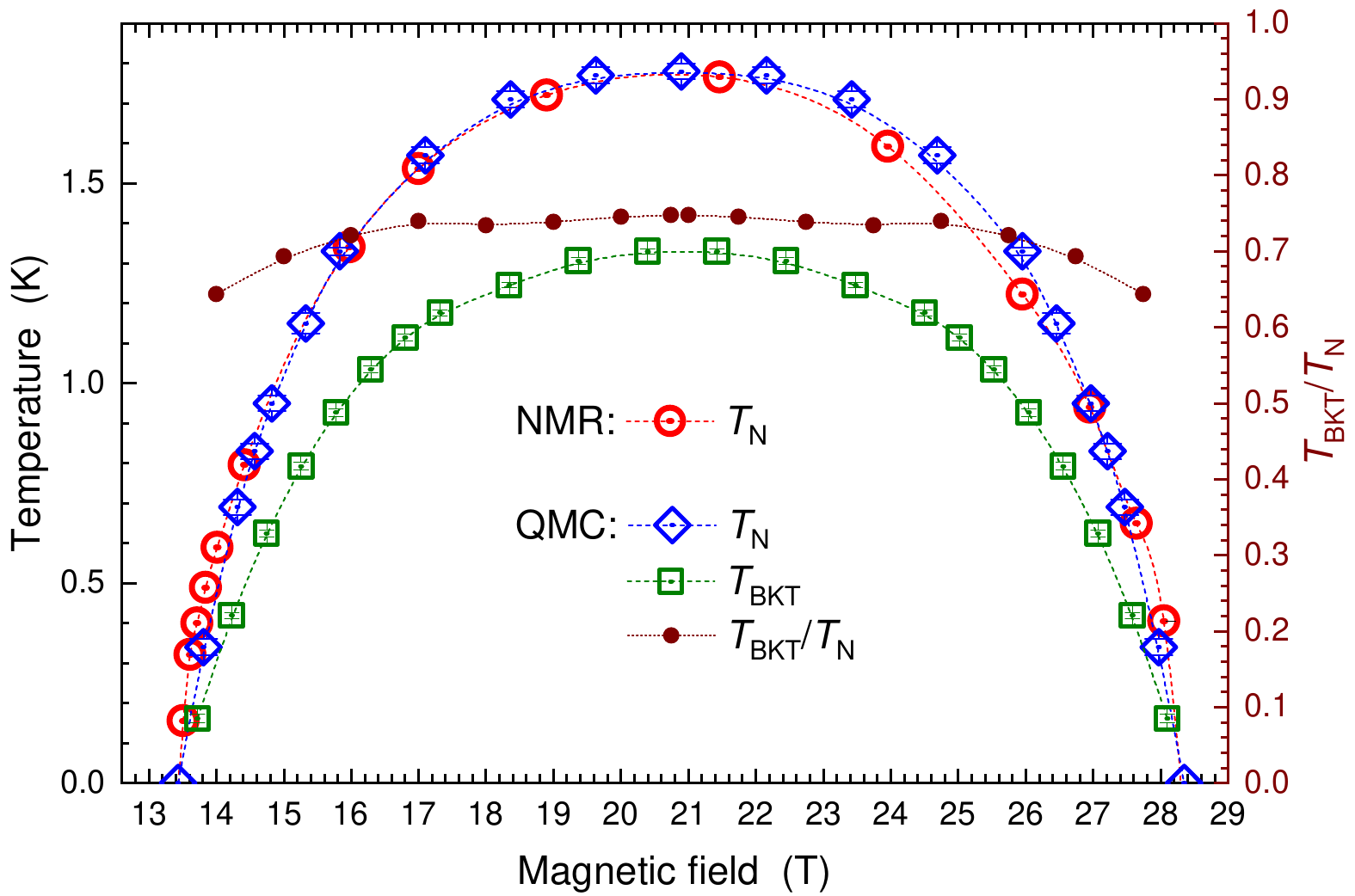}
	\caption{The experimental phase diagram of Ba$_2$CuSi$_2$O$_6$Cl$_2$, obtained by NMR (red circles) from the maximum of critical fluctuations at the phase transition--measured by the peak of the $T_1^{-1}(T)$ or $T_1^{-1}(H)$ data (see Figs.~\ref{Fig_T1vsT} and \ref{Fig_CriticalPeaks}), is compared to the theoretical fit by the QMC simulations (blue diamonds). The latter fit allows for the QMC determination of the $T_{\textrm{BKT}}$ (green squares and brown dots for the $T_{\textrm{BKT}}/T_N$ ratio). Lines connecting the data points are a guide to the eye.}
	\label{Fig_PhaseDiagram}
\end{figure}

\begin{figure}[t!]
	\centering
	\includegraphics[width=0.6\columnwidth]{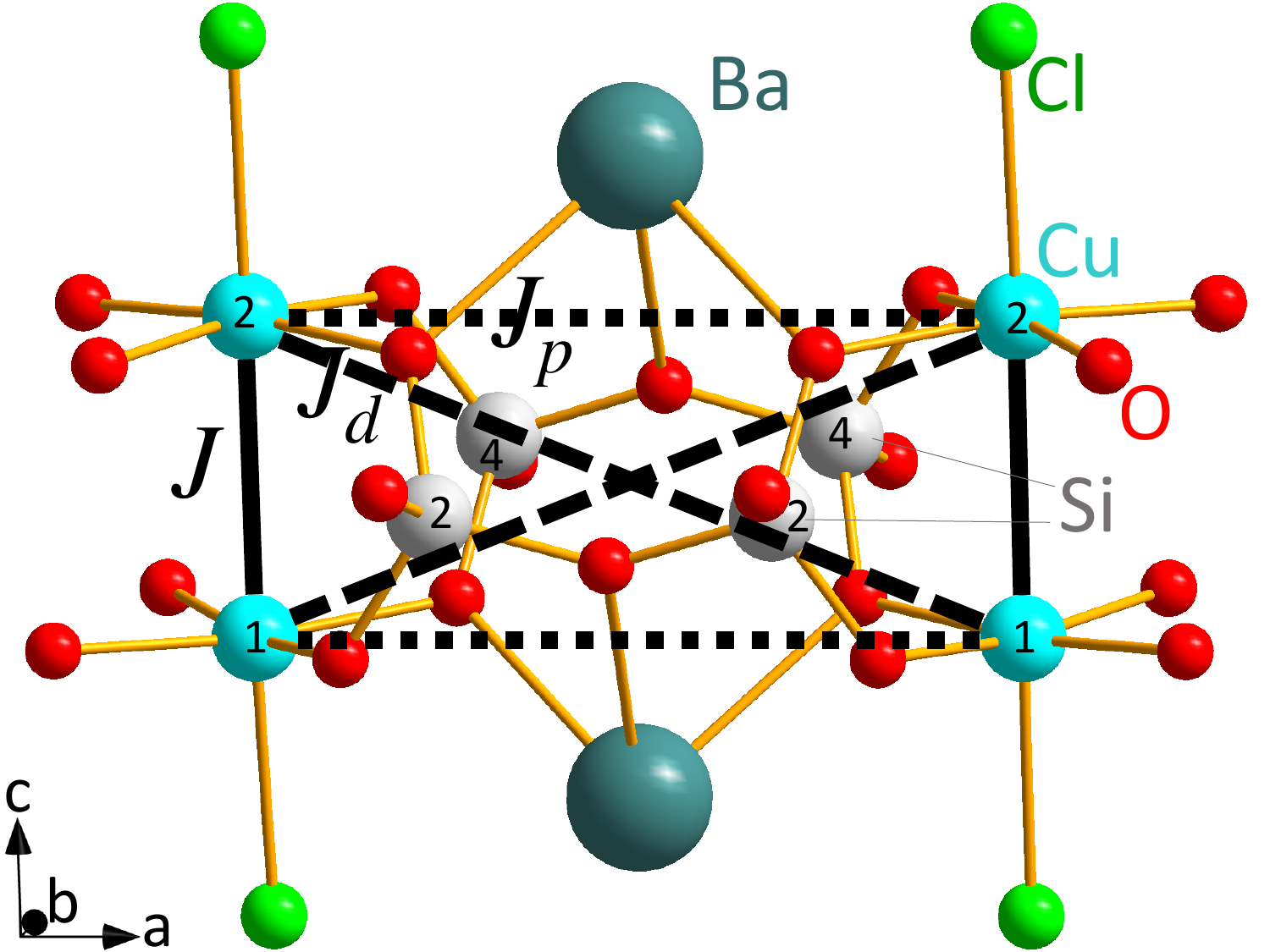}
	\caption{{\color{black}Structural element} of the Ba$_2$CuSi$_2$O$_6$Cl$_2$ crystal structure that contains two neighboring Cu$^{++}$ spin dimers
and provides the main antiferromagnetic exchange couplings $J$, $J_d$ and $J_p$, respectively denoted by the solid, dashed and dotted thick black lines. Numbers on the Cu and Si atoms present different crystallographic sites \cite{Nawa2019}. The exchange coupling paths between the dimers along the $a$ axis passes alternatively either through Si(2)-Si(2) and Si(4)-Si(4) atoms{\color{black}, which is the configuration} shown in the figure{\color{black},} or through Si(1)-Si(1) and Si(3)-Si(3) atoms, while all the couplings along the $b$ axis are identical, passing through Si(1)-Si(2) and Si(4)-Si(3) atoms. {\color{black}The latter two configurations are not shown in the figure.} This provides a network of slightly different $J_d$ and $J_p$ values presented in Ref.~\cite{Nawa2019}. In our theoretical modeling we have neglected these details and considered a simplified, \emph{average} tetragonal (direction-independent) structure.}
	\label{Fig_Couplings}
\end{figure}

In contrast to this, Ba$_2$CuSi$_2$O$_6$Cl$_2$ is a perfectly stoichiometric compound, which has a big advantage that its critical fields are much lower \cite{Okada2016}, 13.4 and 28.4~T for the field ($H$) perpendicular to the planes orientation used in this study (Fig.~\ref{Fig_PhaseDiagram}), which considerably simplifies studies of its phase diagram. While the structure of the 2D planes (Fig.~\ref{Fig_Couplings}) is closely related to those of the Han purple, Ba$_2$CuSi$_2$O$_6$Cl$_2$ is not tetragonal but orthorhombic, whereas the unit cell of a single layer contains two Cu$^{++}$ spin dimers, but, unlike in Han purple, all the dimers are equivalent. The structure of the exchange couplings is a bit involved \cite{Nawa2019} but the departure from the tetragonal symmetry is very small, so for theoretical modeling we retained a simplified tetragonal model: The spins of a dimer have the intradimer coupling $J\approx  28$~K and \emph{all} their interdimer couplings are equivalent and defined by the dominant diagonal coupling $J_d \approx 4$~K, which is slightly frustrated by the parallel coupling $J_p\ll J_d$~\cite{Okada2016}, whereas all the couplings are antiferromagnetic. Note however that there is a significant uncertainty on the actual values of those interdimer couplings $J_{p,d}$ since they were obtained in Ref.~\cite{Okada2016} from fitting the magnetization curve using exact diagonalization data on very small 32-site ($4\times 4\times 2$) cluster. As has been done for the undoped and doped Han purple to simplify QMC simulations \cite{Allenspach2020,Allenspach2021}, this spin-dimer Heisenberg Hamiltonian can be reduced to an effective XXZ spin Hamiltonian~\cite{Totsuka1998,Mila1998,Giamarchi1999} by removing from the original Hilbert space, which has four states per spin-1/2 dimer, the upper two states of each dimer and retaining only the two lower ones. These two states define an effective spin-1/2 or, equivalently, a hard-core (hc) boson. In this latter description, the average coupling defines the hc-boson interaction $V_0 = (J_p + J_d)/2$, while the hopping term $t$ is reduced by frustration $t = (J_p - J_d)/2$ \cite{Abendschein2007}. Finally, one can apply the contractor-renormalization (CORE) correction that improves the energy match between the original and the reduced states \cite{Abendschein2007}, leading to the corrected interaction $V = V_0 - 3t^2/[2(J-V_0)]$.

The most realistic hopping and interaction parameters for the effective hc bosons can be inferred from QMC simulations~\cite{Sandvik2002}, trying to match the theoretical and experimental phase diagrams. Large-scale and low-temperature computations have been performed for 3D coupled layers of sizes
$L\times L\times L/2$, with  $L = 12$, 16, 20, \ldots,
32, where individual layers are governed by the effective spin-$\frac{1}{2}$ XXZ Hamiltonian
\begin{equation}
{\cal{H}}_{\rm 2D}=\sum_{\langle i\,j\rangle}t\left(S_i^{+}S_j^{-} +{\rm{H.c.}}\right)+V S^z_i S^z_j -h\sum_{i}S^z_{i},
\label{eq:H2D}
\end{equation}
and are 3D coupled 
through
\begin{equation}
\label{eq2}
{\cal{H}}_{\color{black}\textrm{3D}}=\sum_{\langle i\,j\rangle_{\color{black}\textrm{3D}}}t_{\color{black}\textrm{3D}}\left(S_i^{+}S_j^{-} +{\rm{H.c.}}\right)+V_{\color{black}\textrm{3D}} S^z_i S^z_j.
\end{equation}
Using the following parameters $t=1.97$~K, $V=1.85$~K, $t_{\color{black}\textrm{3D}}=0.03$~K, and $V_{{\color{black}\textrm{3D}}}=-0.03$~K, we obtain excellent quantitative agreement with the experimentally obtained BEC phase boundary; see Fig.~\ref{Fig_PhaseDiagram}. Note that these $t$ and $V$ values are compatible with the previously given formulas and estimates of the exchange couplings. We also remark that the effective field $h$ is related to the true external magnetic field $H$ by
$h={\color{black}g}\mu_BH-J-2V-V_{{\color{black}\textrm{3D}}}$. Taking $J=28.1$~K and the gyromagnetic factor ${\color{black}g}= 2.26(2)$ allows us to quantitatively agree very satisfactorily with the two experimental estimates of the critical fields: the effective XXZ Hamiltonian yields $H_{c1}=13.3(2)$~T and $H_{c2}=28.5(3)$~T.

Finally, the most sensitive parameter for matching the critical temperature dome is the interlayer coupling, as expected from the very strong logarithmic enhancement of the 3D critical temperature $T_N$ with $t_{\color{black}\textrm{3D}}$~\cite{Laflorencie2012}. $T_N$ is extracted from QMC simulations, at each applied field value, using a standard finite-size scaling analysis~\cite{Sandvik1998} of the superfluid stiffness $\rho_{\rm sf}$, which scales as $L^{2-D-z}$ at criticality (here we fix dimension $D=3$ and the dynamical exponent of the finite temperature transition $z=0$). Given the approximations used, the BEC phase boundary obtained in this way is remarkably close to the experimental one, and presents only minor deviations (Fig.~\ref{Fig_PhaseDiagram}). In particular, in contrast to the original spin-dimer Hamiltonian, its reduced hc-boson version is perfectly symmetric with respect to the middle of the BEC dome and, therefore, cannot account for the slight asymmetry of the experimental phase boundary. Furthermore, the experimental $T_N(H)$ dependence approaches to its edges (the quantum critical points) somewhat steeper than the QMC simulation. This is probably related to some subdominant terms of the Hamiltonian{\color{black}, which are unknown and thus cannot be modeled. We suspect that they exist because our preliminary very low temperature $T_{1}^{-1}$ NMR data taken at the two critical fields seem to present an abrupt change of spin dynamics at very low temperature, below 0.15~K at $H_{c1}$ and below 0.4~K at $H_{c2}$. We speculate that this might indicate the presence of some unexpected very low temperature phases, but did not pursue our investigation in this direction.}

In view of remarkable agreement between the experimental and modeled phase boundary, we can use the same model to predict the corresponding BKT \cite{BKT} transition temperature $T_{\textrm{BKT}}$ that the system would present if the dimer planes were perfectly decoupled, true 2D systems. This parameter is essential to properly define and constraint the usage of a purely 2D description to describe the spin dynamics in real compounds, which are necessarily quasi-2D.
For the effective 2D XXZ model (with $t_{\color{black}\textrm{3D}}=V_{\color{black}\textrm{3D}}=0$), the critical temperature $T_{\textrm{BKT}}$ can be obtained using the Nelson-Kosterlitz relation~\cite{Nelson1977} $\rho_{\rm sf}(T_{\textrm{BKT}})=2T_{\textrm{BKT}}/\pi$. However, this is only true in the thermodynamic limit and for finite size data, non-negligible logarithmic
corrections are expected for BKT transitions~\cite{Weber1987}. Given $T^*(L)$, the finite-size
solution of $\rho_{\rm sf}(L)=2T/\pi$, the thermodynamic limit estimate of
$T_{\textrm{BKT}}$ is obtained from the finite-size expression~\cite{Bramwell1993} $T^*(L)=T_{\textrm{BKT}}+C/(\ln L)^2$ valid for large $L$, where $C$ is a nonuniversal constant. It is interesting to note that the $T_{\textrm{BKT}}/T_N$ ratio is nearly field independent over the central half of the BEC dome, where it amounts to 0.74, and  decreases only weakly toward the edges.

\section{NMR relaxation rate}
\subsection{Experimental results}

NMR experiments were performed on a Ba$_2$CuSi$_2$O$_6$Cl$_2$ single-crystal sample (4$\times$1.9$\times$0.35 mm$^3$), synthesized as described in Ref.~\cite{Okada2016}. The sample was oriented to have the $c$ axis parallel to the applied magnetic field $\bm{H}${\color{black}, with an error of less than 2$^{\circ}$.} For magnetic field values up to 18.9~T we used superconductive magnets, equipped either with a standard cold-bore variable-temperature insert (VTI) for temperatures above 1.4~K or a $^3$He-$^4$He dilution refrigerator for the measurements below 1~K, where the sample was placed inside the mixing chamber. The higher magnetic field values ($\geq 18.9$~T) were obtained by a M9 resistive magnet at LNCMI, where the $^3$He cryostat was used to attain temperatures down to 0.4~K.

\begin{figure}
	\centering
	\includegraphics[width=\columnwidth]{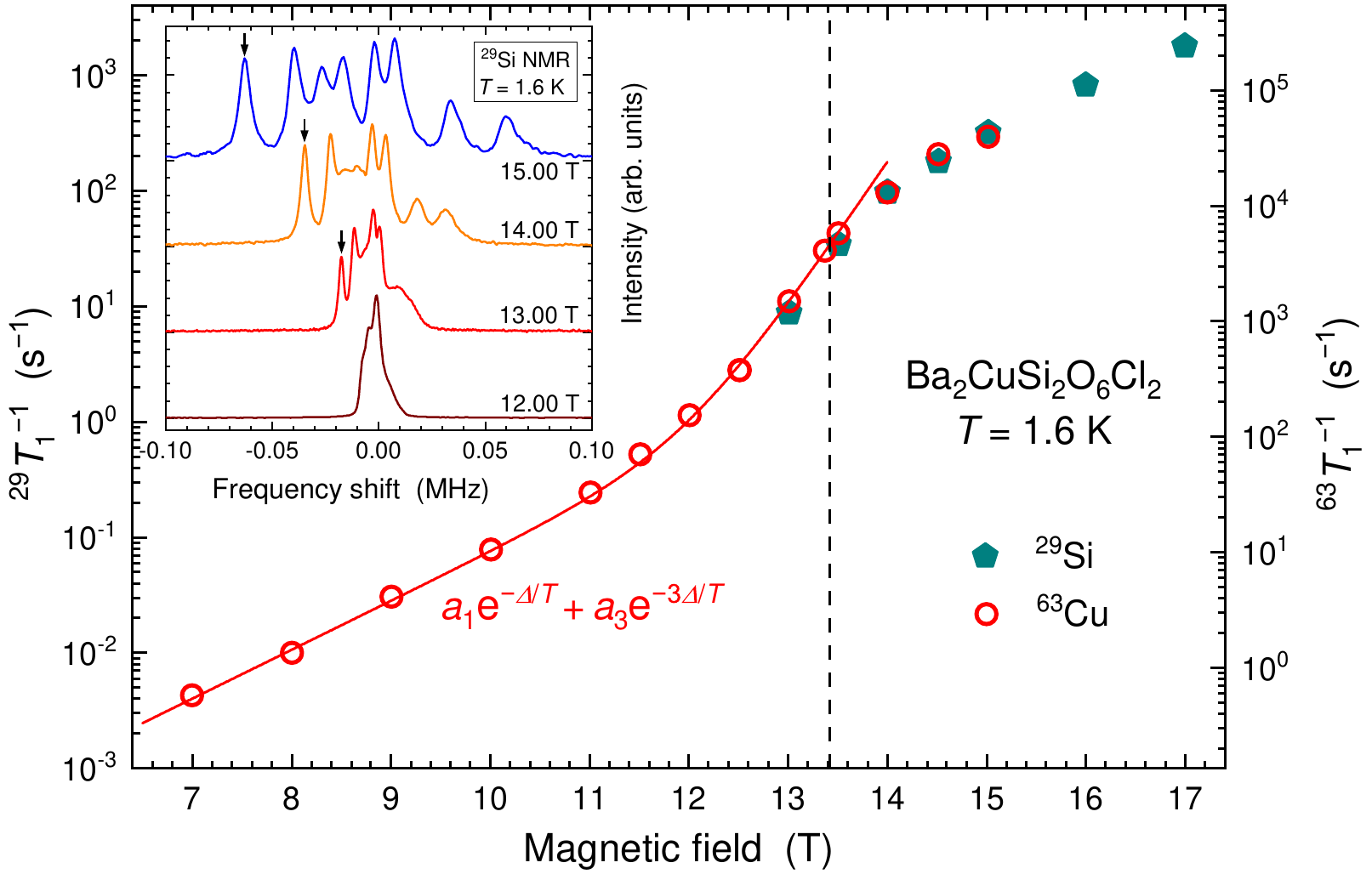}
	\caption{Magnetic field dependence of nuclear spin-lattice relaxation rate $T_1^{-1}$ measured by $^{63}$Cu (red circles) and $^{29}$Si (dark cyan pentagons) NMR at 1.6~K. Solid line is a two-parameter (prefactors) two-exponential fit that confirms the opening of the Zeeman gap $\Delta$ below the first critical field $\mu_0 H_{c1}=13.42~T$, and the presence of the critical fluctuations near $H_{c1}$, characterized by three-magnon processes \cite{Ranjith2022}. Inset shows the field dependence of the $^{29}$Si NMR spectra, where arrows denote the low-frequency peak used for the $T_1^{-1}$ measurements. At low field and temperature values, where the spin polarization is nearly zero and the $^{29}$Si NMR spectrum is thus unresolved, $T_1^{-1}$ was measured by $^{63}$Cu NMR. The two vertical scales are related by $^{63}T_1^{-1}/^{29}T_1^{-1}=135$.}
	\label{Fig_GapedCriticalT1}
\end{figure}

\begin{figure}
	\centering
	\includegraphics[width=\columnwidth]{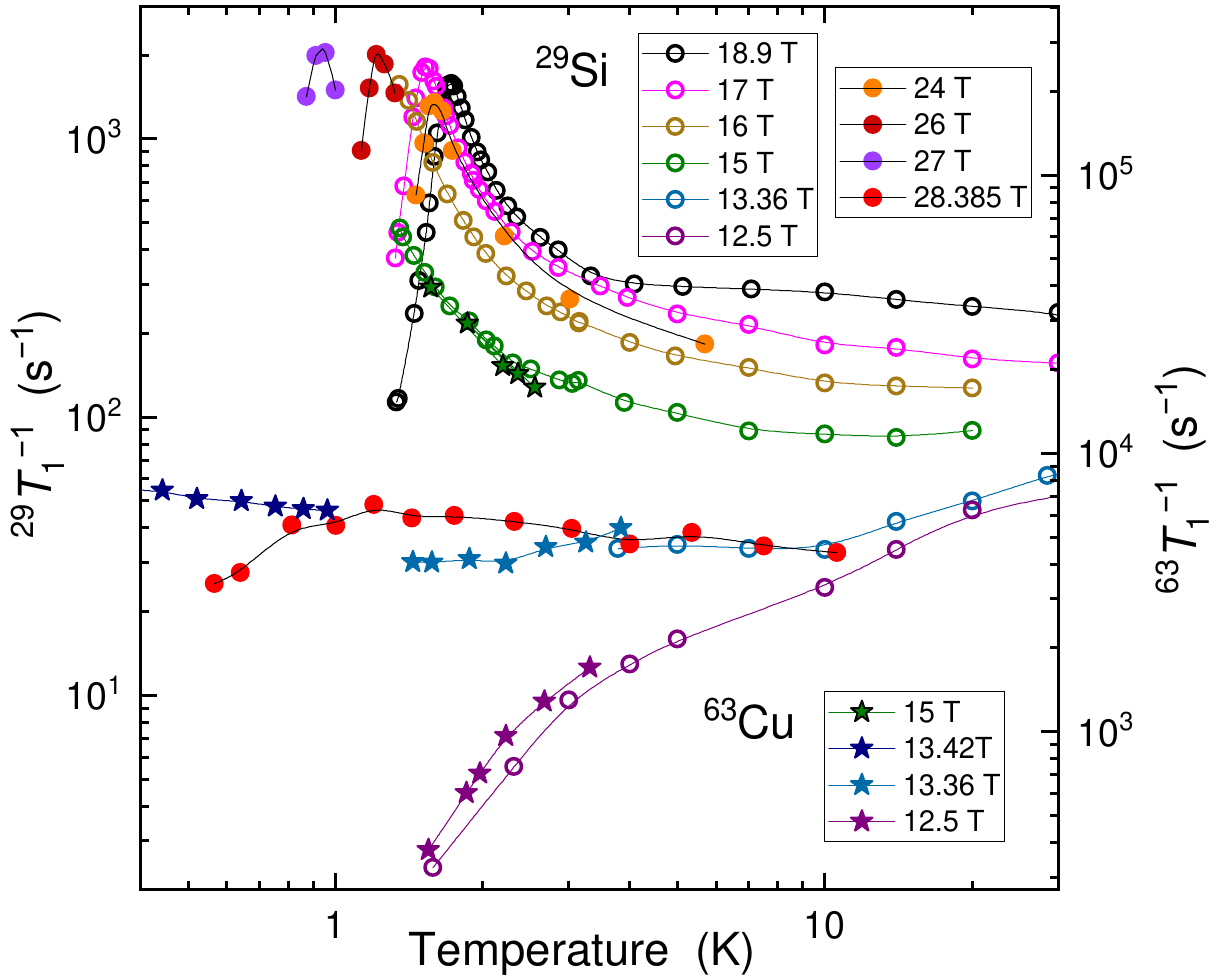}
	\caption{Temperature dependence of the $^{29}$Si (circles) and $^{63}$Cu (stars) $T_1^{-1}$ NMR data for fixed magnetic field values covering the whole phase diagram. At the phase transition into the low-temperature BEC phase, the data sets present very strong peak that reflects the critical spin fluctuations. At both critical fields, $H_{c1}$ and $\mu_0 H_{c2}=28.385~T$, the relaxation rate is nearly constant, while below $H_{c1}$ it reflects the gap opening. The two vertical scales are related by $^{63}T_1^{-1}/^{29}T_1^{-1}=135$. Lines connecting the data points are a guide to the eye.}
	\label{Fig_T1vsT}
\end{figure}

In this article we focus on the low-energy spin dynamics as observed through the nuclear spin-lattice relaxation rate $T_1^{-1}$ of $^{63}$Cu, which are the on-site nuclei, and $^{29}$Si, which are neighboring nuclei that are coupled to both electronic spins of a dimer. The detailed structural data \cite{Nawa2019} distinguish two Cu sites in each dimer (Fig.~\ref{Fig_Couplings}) and, indeed, in the $^{63,65}$Cu NMR spectrum we observe two sets of lines belonging to two Cu sites having slightly different quadrupolar couplings $^{63}\nu_Q$~= 31.43 and 29.72~MHz and slightly different hyperfine shifts \cite{Ranjith_unpublished}. For technical convenience, the $^{63}T_1^{-1}$ data have been measured on the high-frequency satellite of the $^{63}$Cu spectrum that has higher quadrupolar coupling. {\color{black}The $^{29}$Si NMR spectra show eight lines (see Appendix~\ref{appB})} which spread proportionally to the magnetic field-induced or/and temperature-induced spin polarization (inset in Fig.~\ref{Fig_GapedCriticalT1}). The $^{29}T_1^{-1}$ data have been taken on the lowest frequency line as soon as this line was resolved in the spectrum. When the spin polarization is close to zero this is not the case, so $^{63}T_1^{-1}$ data were used instead. Overlapping the $^{29}T_1^{-1}$ and the $^{63}T_1^{-1}$ data in Figs.~\ref{Fig_GapedCriticalT1} and \ref{Fig_T1vsT} shows the $^{63}T_1^{-1}$ rate is 135 times faster than $^{29}T_1^{-1}$, which is the ratio of the two vertical axes of the two figures.

\begin{figure}
	\centering
	\includegraphics[width=\columnwidth]{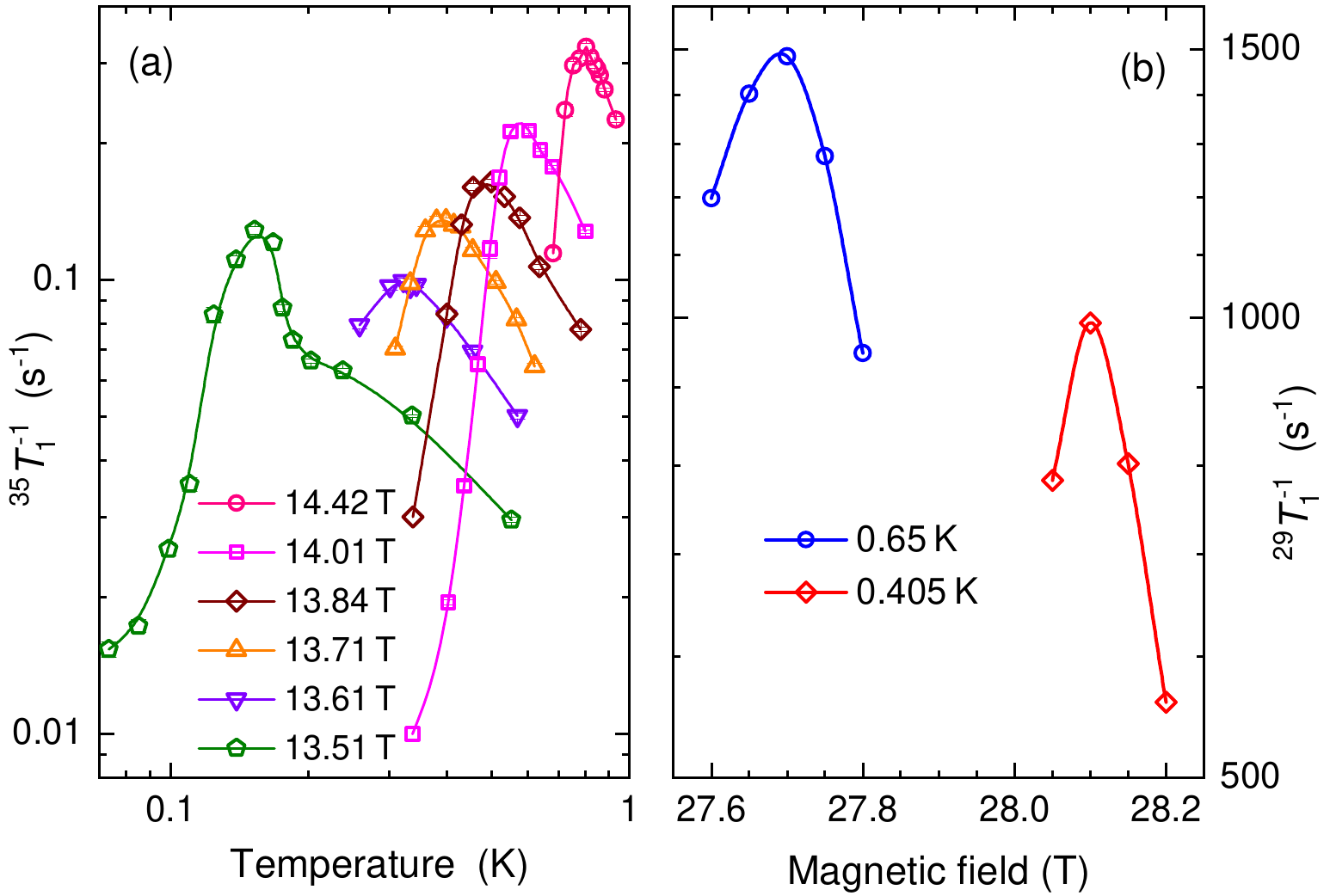}
	\caption{{\color{black}(a) Temperature dependence of the $^{35}$Cl $^{35}T_1^{-1}$ at fixed field values and (b) magnetic field dependence of the $^{29}$Si $^{29}T_1^{-1}$ at fixed temperatures, recorded close to their peak values in order to define the low-temperature edges of the BEC phase boundary. Lines are the B-spline (a) and the modified Bézier (b) interpolations, used to precisely define the peak position.}}
	\label{Fig_CriticalPeaks}
\end{figure}

The $T_1^{-1}$ rate was measured by the saturation-recovery method, and the time ($t$) recovery of the nuclear magnetization $M(t)$ after a saturation pulse was fitted by an exponential function $\propto$\,$\exp(-t/T_1)$ for the $^{29}$Si $I = 1/2$ nuclei and by the three-exponential function $\propto$\,$[0.1\exp(-t/T_1)+0.5\exp(-3t/T_1)+0.4\exp(-6t/T_1)]$ for the satellite line of the $^{63}$Cu $I = 3/2$ nuclei and relaxation of magnetic origin. {\color{black}In Fig.~\ref{Fig_CriticalPeaks}(a) we also show $^{35}T_1^{-1}$ data recorded on the highest intensity central line of the spin-3/2 $^{35}$Cl nuclei and fitted by the corresponding two-exponential function $\propto$\,$[0.1\exp(-t/T_1)+0.9\exp(-6t/T_1)]$.}
The agreement of the measured $M(t)$ curves and the theoretically expected fits confirms that the sample is homogeneous, without any noticeable distribution of the local $T_1^{-1}$ values. This is indeed the case for nearly all our data, the exception being the regimes where inhomogeneous $T_1^{-1}$ distributions could be expected. One example is the very low-$T$ end ($T \leq 1.0$~K) of the $^{29}T_1^{-1}$ taken at the upper critical field $H_{c2}$ (Fig.~\ref{Fig_T1vsT}), as these data points are taken very close to the phase transition line and strongly capture any distribution of the local $H_{c2}$ values over the sample. Another case is the gapped regime, where the low-frequency line of the $^{29}$Si spectra taken at 12.5~T and below 6~K is not properly resolved (inset in Fig.~\ref{Fig_GapedCriticalT1}), so the measurement captures a mixture of different contributions. In these cases the exponential functions in the fits are replaced by the stretched exponential $\exp[-(t/T_1)^{\beta}]$, where the stretching exponent $\beta$ provides a measure of the distribution of the local $T_1^{-1}$ values over the sample \cite{Johnston2006,Mitrovic2008}. In the above mentioned cases, the $\beta$ values at the low-$T$ end decrease down to 0.7, corresponding to moderately inhomogeneous $T_1^{-1}$ values, where the fitted (``average'') $T_1^{-1}$ value is rather insensitive to the presence of the distribution.

Figure~\ref{Fig_GapedCriticalT1} presents the magnetic field dependence of the $^{63}$Cu and $^{29}$Si $T_1^{-1}$ relaxation rates, recorded at 1.6~K. As the logarithmic scale converts an exponential function into its exponent, the apparent linear dependence  below 11~T points to a linear gap opening. The slope of this dependence corresponds precisely to the opening of the single-magnon Zeeman gap: $\Delta(H) = g_c\mu_B\mu_0(H-H_{c1})/k_B$ (here in kelvin units), where $g_c = 2.32$ is the {\color{black}$c$-axis} $g$ factor \cite{Okada2016} and $\mu_{\rm B}/k_{\rm B}$~= 0.67171~K/T, as expected below the critical field $H_{c1}$. Close to $H_{c1}$ we observe a triple-gap (three-magnon) contribution, expected for the critical excitations described by the second order terms in Ref.~\cite{Orignac2007}. Altogether, we successfully fit the whole $^{63}T_1^{-1}(H)$ dependence between 7 and 13.5~T by the two-exponential fit $a_1 e^{-\Delta(H)/T}+a_3 e^{-3\Delta(H)/T}$ that has only two parameters, namely the exponential prefactors $a_1$ and $a_3$. This is a recognized signature of the gapped and the critical regime associated to a single-magnon BEC condensation, see \cite{Ranjith2022} and references therein.

Figure~\ref{Fig_T1vsT} presents an overview of the temperature dependence of $T_1^{-1}$ relaxation rates recorded at fixed magnetic field values covering the whole phase diagram, above and at the phase transition into the BEC phase. The most notable feature is the big peak in the $T_1^{-1}(T)$ dependence that directly reflects the critical spin fluctuations at the phase transition. The maximum of the peaks shown in this figure defines the experimental phase transition temperatures $T_N$ shown in Fig.~\ref{Fig_PhaseDiagram} for the points above 1.3~K. {\color{black}Figure~\ref{Fig_CriticalPeaks}(a) presents the analogous data recorded in the dilution refrigerator below 15~T, used to define the phase boundary on its low field $(H_{c1}$) edge, and Figure~\ref{Fig_CriticalPeaks}(b) shows the magnetic field dependence $T_1^{-1}(H)$ taken at constant temperatures, used to define the high field $(H_{c2}$) edge of the BEC dome.} Extrapolating these two $T_N(H)$ edges to the zero temperature by a theoretically expected critical dependence $T_N(H) \propto |H - H_c|^{2/3}$ we define the experimental values for the two critical fields, $\mu_0 H_{c1}$~= 13.42~T and $\mu_0 H_{c2}$~= 28.385~T, in agreement with the previous determination from the magnetization data \cite{Okada2016}.

With regard to the shape of the $T_1^{-1}(T)$ peaks, we note that their relative size (base to maximum) reaches up to one order of magnitude, which is much larger than what is expected and observed in the quasi-1D systems \cite{Horvatic2020}. The peak extends typically well beyond 2$T_N$, which is definitely much further than in the 3D systems, where the critical fluctuations are localized within $\approx \pm$20\% of $T_N$. This is also clearly less than the extension of the purely 1D, power-law dependence observed in quasi-1D systems \cite{Horvatic2020}. Altogether, the very shape of the $T_1^{-1}(T)$ peak, as compared to what is observed in quasi-1D and in 3D systems provides a clear signature, from the spin dynamics, of the 2D character of the system.

At the two critical fields, we observe a very weak temperature dependence. Close to the critical fields, this dependence is \emph{very} sensitive to the chosen field value and presents in general a fan-shaped plots \cite{Mukhopadhyay2012} as the magnetic field is varied from the gapless regime above the BEC phase across the quantum critical field value into the gapped regime. As much as we could precisely determine the two $H_{c1}$ and $H_{c2}$ values and obtain the corresponding reliable $T_1^{-1}(T)$ dependences (which proved to be technically challenging at $H_{c1}$), we find identical values and temperature dependences. Using the most reliable data, $^{29}T_1^{-1}$ taken at $H_{c2}$, we find that $T_1^{-1}(T) \propto T^{-0.14}$ in the 1-10~K temperature range. We remark that this compares very favorably to the predicted quantum critical behavior at the magnetic instability in AFM 2D metals,
$T_1^{-1}(T)$~$\propto -\ln T$ \cite{Moriya2003}, but it is not evident to what extent this is applicable to an insulating spin system.

Notably, both the $T_1^{-1}(T)$ dependence and the absolute values of the relaxation rates are the same for both critical fields. This is different from what is expected and observed in quasi-1D compounds: The fully polarized state above $H_{c2}$ is a classical one, while the zero-polarization state below $H_{c1}$ is a complicated superposition of spin singlets that is of purely quantum origin. When describing physical quantities in this latter state {\color{black}of quasi-1D compounds,} we have to apply the renormalization induced by quantum fluctuations \cite{Kohama2011} that are absent above $H_{c2}$. This has been successfully applied in the quasi-1D compound NiCl$_2$-4SC(NH$_2$)$_2$ (DTN) \cite{Kohama2011}, but the corresponding effect is absent from our $T_1^{-1}$ data in a quasi-2D Ba$_2$CuSi$_2$O$_6$Cl$_2$. {\color{black}This suggests that the renormalization effects are negligible in the quasi-2D case, possibly due to diminished strength of the 2D fluctuation as compared to the 1D ones.}
We remark that the reduced hc-boson Hamiltonian is \emph{symmetric} with respect to the center of the BEC dome [$(H_{c1}+H_{c2})/2$], which is the zero of the corresponding \emph{effective} field. In this representation, both critical fields are the effective saturation fields, and their behavior is thus necessarily identical. So, only by considering the full spin-dimer Hamiltonian we can account for the asymmetry between the two critical fields. Why would {\color{black}this asymmetry be apparent in the $T_1^{-1}$ data of quasi-1D compounds and not in quasi-2D compounds remains an open question} to be investigated theoretically.  However, it is also possible that the equality of $T_1^{-1}$ at both critical field in Ba$_2$CuSi$_2$O$_6$Cl$_2$ is a result of some kind of compensation, be it accidental or not.

Finally, below $H_{c1}$, at 12.5~T, as the temperature is lowered, we observe a strong decrease of spin fluctuations induced by the presence of a gap that we discussed in relation to the Fig.~\ref{Fig_GapedCriticalT1}.


\subsection{QMC simulations}

In a system dominated {\color{black}by} magnetic fluctuations, the nuclear spin-lattice relaxation of a given nuclear site originates from the fluctuations of the transverse component of the local field ${\mathbf{h}}$ experienced by these nuclei, induced by the fluctuations of the neighboring electronic spins ${\mathbf{S}}$, whereas the two quantities are linearly related by the hyperfine coupling tensor ${\mathbf{A}}$,  ${\mathbf{h}}(t) = - {\mathbf{A}}{\mathbf{S}}(t)$. The spin fluctuations are described by the dynamic correlation function (where the average spin values are taken to be zero)
\begin{equation}
\label{eq3}
S_{\mathbf{q}}^{\mu \nu}(\omega)= \int_{-\infty}^{+\infty}\left< S_{\mathbf{-q}}^{\mu}(t) S_{\mathbf{q}}^{\nu}(0) \right> e^{-i \omega t} dt ,
\end{equation}
where $\mu$, $\nu$ denote the coordinate axes $X, Y, Z$ and $\left< \right>$ the thermal average. In the external magnetic field applied along the $z$ direction, the $T_1^{-1}$ rate is given by the Moriya's formula \cite{Moriya1956}
\begin{equation}
\label{eq4}
T_{1,Z}^{-1} = \frac{\gamma^2}{2} \sum_{\mathbf{q}} \sum_{\mu = X,Y,Z} [A_{X\mu}(\mathbf{q})^2 + A_{Y\mu}(\mathbf{q})^2] S_{\mathbf{q}}^{\mu \mu}(\omega_0) ,
\end{equation}
where $\gamma$ is the gyromagnetic ratio of the nucleus, $S_{\mathbf{q}}^{\mu \nu}$ is considered to be a diagonal tensor, and $\omega_0$ is the NMR frequency. The latter is typically the smallest energy scale, of the order of 0.01~K in our case, and can be thus replaced by zero. In general, both transverse and longitudinal fluctuations ($S_{\mathbf{q}}^{+-}$ and $S_{\mathbf{q}}^{ZZ}$) contribute, and in the sum over $\mathbf{q}$ they are filtered by the form-factors reflecting the $\mathbf{q}$-dependence of the hyperfine coupling. In our case, the dominant spin fluctuations are the transverse ones, as they lead to a transverse order parameter in a BEC phase. Furthermore, for the Cu nuclei the on-site coupling is dominant, leading to the $\mathbf{q}$-independent ${\mathbf{A}}$. Altogether, we arrive to the simplest expression
\begin{eqnarray}
\nonumber
T_{1}^{-1} = \sum_{\mathbf{q}} T_{1}^{-1}({\mathbf{q}}) , \textrm{~where}  \\
\label{eq5}
T_{1}^{-1}({\mathbf{q}}) \propto [S_{\mathbf{q}}^{+-}(0) + S_{\mathbf{q}}^{-+}(0)]/2  ,
\end{eqnarray}
which is what has been calculated by the QMC simulations for the previously defined model, using the procedure defined in Ref.~\cite{Dupont2018}, to compare the measured $T_1^{-1}(T)$ data to theoretical predictions. For the Si nuclei, we might in principle expect some effects of form factors and some weak contribution of longitudinal fluctuations, which are coupled by the off-diagonal elements of the ${\mathbf{A}}$ tensor. However, from the overlap of the $^{63}T_1^{-1}$ and $^{29}T_1^{-1}$ data in Figs.~\ref{Fig_GapedCriticalT1} and \ref{Fig_T1vsT}, we conclude that these effects are negligible, so Eq.~(\ref{eq5}) is thus used to describe all the $T_1^{-1}$ data.

Computing $S^{\mu\nu}_{\bf q}(\omega)$ requires analytic continuation of the QDM data obtained in imaginary time. We have used stochastic analytic continuation (SAC)~\cite{SAC} and refer to Ref.~\cite{Dupont2018} for technical details.

\begin{figure}[t!]
	\centering
	\includegraphics[width=\columnwidth]{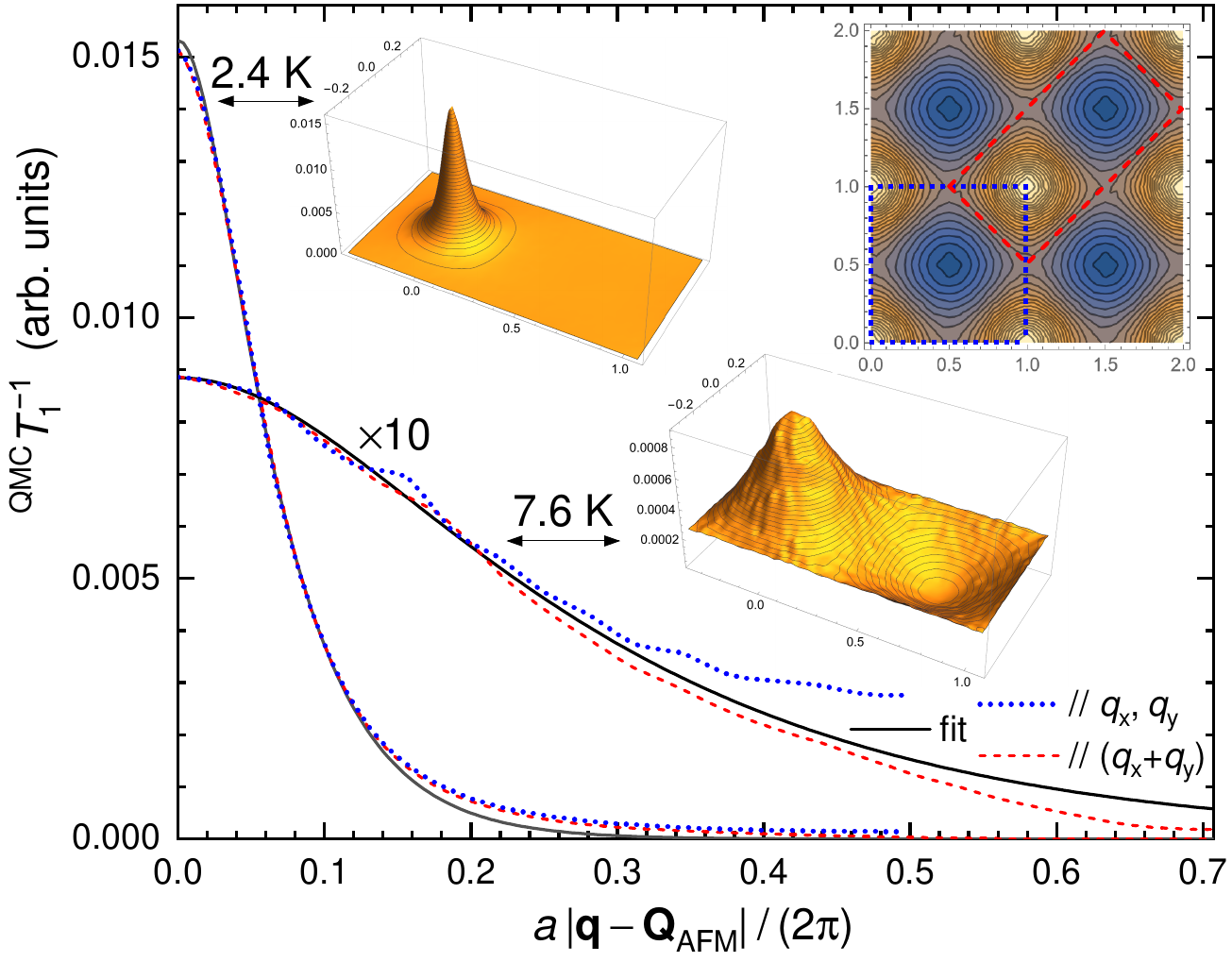}
	\caption{The wave-vector dependence $T_1^{-1}(\mathbf{q})$ of the QMC simulations for 17 T presents a peak at the antiferromagnetic wave vector $\mathbf{Q}_{\rm{AFM}}$. For the two characteristic temperatures (2.4 and 7.6~K), the blue dotted line and the red dashed line present the $T_1^{-1}(\mathbf{q})$ dependence along the axes ($q_x$, $q_y$) of the unit cell and along its diagonal, respectively. The solid black line is the ``flat-top exponential'' fit of the $T_1^{-1}(\mathbf{q})$ peak, used to define the corresponding correlation length, as explained in the text. The two 3D-plot insets display the complete $\mathbf{q}$-dependence over a conveniently defined unit cell, shown by the red dashed line in the contour plot (for 7.6~K) given in the top-right inset.}
	\label{Fig_T1vsQ}
\end{figure}

To control the effects of dimensionality and the finite size, the QMC determination of $T_1^{-1}$ has been carried out for the 32$\times$32 and 64$\times$64 purely 2D systems and for the 32$\times$32$\times$8 full 3D system. In Fig.~\ref{Fig_T1vsQ} we first examine the obtained
$T_{1}^{-1}({\mathbf{q}})$ dependence, here for the 2D 32$\times$32 simulation at 17 T. As expected for an AFM system, there is a peak at the AFM wave vector ${\mathbf{Q}}_{\rm{AFM}}$, and in the figure we thus use the reduced $a({\mathbf{q}}-{\mathbf{Q}}_{\rm{AFM}})/(2\pi)$ scale, where $a$ is the unit cell size. The contour-plot inset presents four unit cells, to show our nonstandard choice of the unit cell, delimited by red dashed lines, which enables us to capture both the whole (unsplit) peak and the whole zone of the minimum, in contrast to what presents a standard unit cell, delimited here by the blue dotted lines. For the former unit cell, the two 3D-plot insets present the full $T_{1}^{-1}({\mathbf{q}})$ dependence at 2.4 and 7.6~K. We see that the AFM peak strongly grows and narrows on decreasing temperature. It is nearly perfectly axially symmetric, except near the (rectangular) edges of our unit cell when these values are significant (non zero). The mainframe of the figure presents in detail the shape of the AFM peak taken along two representative directions, along the axis and along the diagonal. It turns out that the major upper part of this shape is remarkably well fitted by a ``flat-top exponential'' function
\begin{equation}
\label{eq6}
T_{1}^{-1}({\mathbf{q}}) = a_0 \exp[-\sqrt{(\xi a \Delta{\mathbf{q}})^2 + d^2} + d] ,
\end{equation}
where $\Delta {\mathbf{q}} = {\mathbf{q}}-{\mathbf{Q}}_{\rm{AFM}}$, $a_0$ is the amplitude, $\xi$ describes the exponential decrease and $d$ models the flat-top maximum of the peak. The corresponding three-parameter fits of the QMC $T_{1}^{-1}({\mathbf{q}})$ data show that $d$ is nearly temperature independent, so we could fix its value to $d$~= 0.82 and thus reduce fitting to only two parameters, $a_0$ and $\xi$. For 2D simulations, we have compared thus obtained ``dynamical'' $\xi$ values to the standard ``static'' $\xi$ values obtained independently from the \emph{static} transverse structure factor, which is an equal time correlator
\begin{equation}
\label{eq7}
S_0^{\perp}({\mathbf{q}}) =  \left< S_{\mathbf{-q}}^{+}(t) S_{\mathbf{q}}^{-}(t) \right>,
\end{equation}
where, by virtue of time invariance, we put $t = 0$.
Here, the $\xi$ is defined as the coefficient of the quadratic dependence $S_0^{\perp}({\mathbf{q}}) \propto 1 - (\xi a \Delta {\mathbf{q}})^2$  in the close vicinity of $\Delta {\mathbf{q}} = 0$; see Appendix~\ref{appA}. Figure~\ref{Fig_XvsT} shows that the $\xi$ values obtained in these two very different ways scale remarkably well, with a proportionality factor of $r_{\xi}=0.575$. In Appendix~\ref{appC} we show that this factor can be understood as a simple consequence of scaling arguments. We remark here that the two methods exploit different $\Delta {\mathbf{q}}$ values, the intermediate range for Eq.~(\ref{eq6}) and the zero limit for Eq.~(\ref{eq7}). The latter method has the advantage of accessing to bigger system sizes, up to 128$\times$128, which in turn minimizes or eliminates the size effects.

\begin{figure}
	\centering
	\includegraphics[width=\columnwidth]{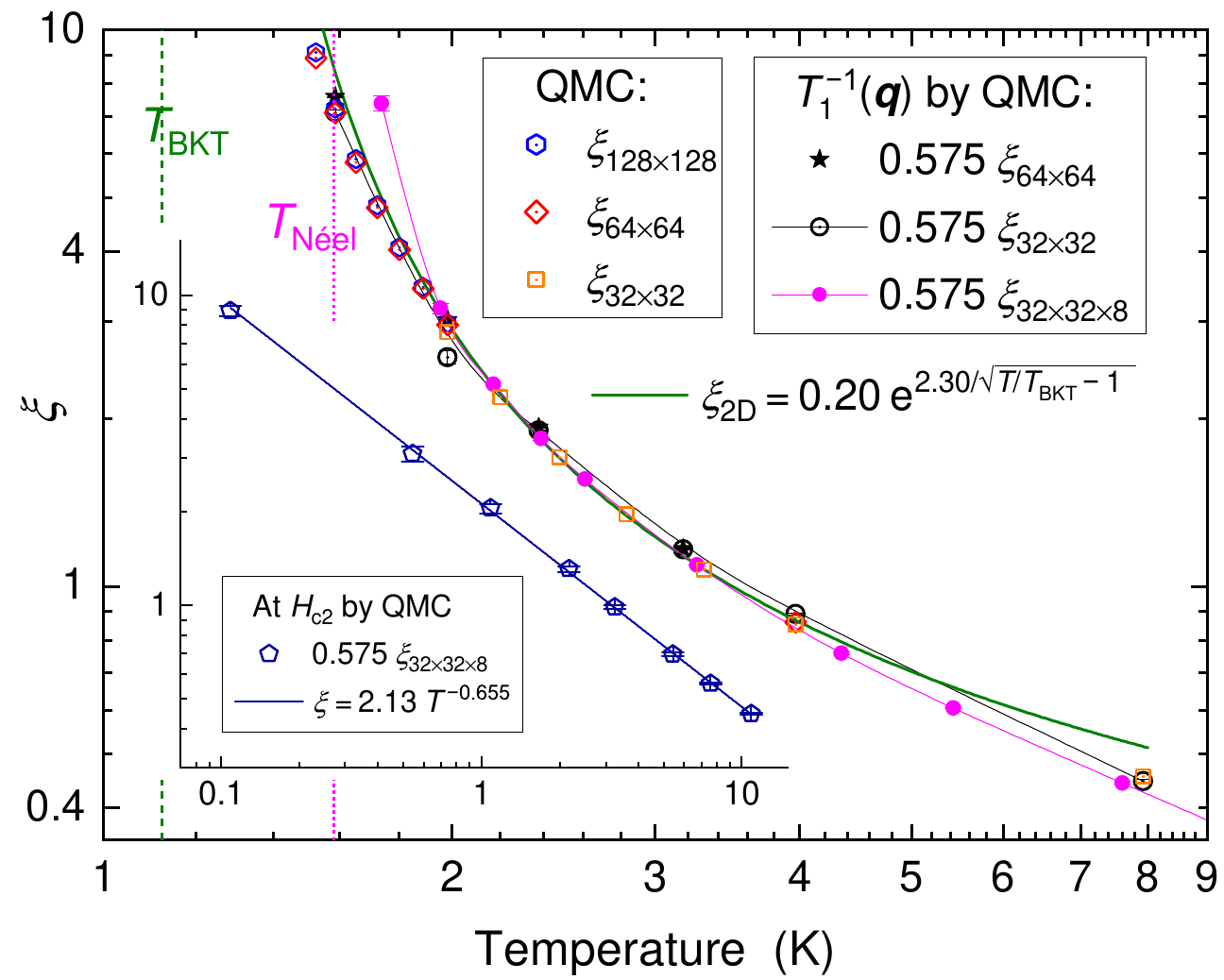}
	\caption{The temperature dependence of the correlation length $\xi$ at equivalently 17 and 24.6 T, determined by QMC either directly or from the fits to the $T_1^{-1}(\mathbf{q})$ peak, for different 2D system sizes as given in the index of $\xi_{L \times L}$. The analytical 2D expression for the 2D-critical $\xi_{\textrm{2D}}$ (green line) asymptotically approaches the QMC data only in the limited range from 1.6\,$T_{\textrm{BKT}}$ to 3\,$T_{\textrm{BKT}}$. The $\xi_{32 \times 32 \times 8}$ values determined from $T_1^{-1}(\mathbf{q})$ for a 3D system differ from the 2D values only very close to $T_N$. Lines connecting the data points are the spline interpolations to guide the eye. {\color{black}The inset shows the $\xi_{32 \times 32 \times 8}$ data points obtained at the critical field $H_{c2}$ and the corresponding power-law fit (solid straight line).}}
	\label{Fig_XvsT}
\end{figure}

In Fig.~\ref{Fig_XvsT}, we have also plotted the theoretically predicted 2D-critical temperature dependence of $\xi$ \cite{Kosterlitz1974,Ding1990}
\begin{equation}
\label{eq8}
\xi_{\textrm{2D}}(T) = \xi_0 \exp(b/\sqrt{T/T_{\textrm{BKT}}-1}),
\end{equation}
where we have fixed the $T_{\textrm{BKT}}$ value to the one calculated by QMC as given in Fig.~\ref{Fig_PhaseDiagram} and used $b$~=2.30 value determined by QMC; see Appendix~\ref{appA}{\color{black}. T}he $\xi_{\textrm{2D}}(T)$ curve approaches asymptotically the numerical values in the $\xi$~$\approx$ 1--10 range, where slight variation of the $b$ and $T_{\textrm{BKT}}$ parameters (compensated by a suitably adjusted prefactor $\xi_0$) only slightly shift this range. In any case, for the $\xi$ values below 1, \textit{a priori}, we do not expect that $\xi_{\textrm{2D}}$ properly describes a particular system. Indeed, for our system, the $\xi_{\textrm{2D}}(T)$ curve departs above the calculated $\xi$ values below $\xi\approx$ 1.0, which corresponds to temperatures above 3.5~K $\approx$ 3.1$T_{\textrm{BKT}}$ $\approx$ 2.2\,$T_N$. This precisely defines the high-temperature validity range for any description based on the anticipated $\xi_{\textrm{2D}}(T)$ given by Eq.~(\ref{eq8}).

On the low temperature side, in practice, the applicability of a purely 2D description will be limited by the presence of a 3D ordering at $T_N$. Knowing that the corresponding 3D fluctuations extend up to $\approx$1.2\,$T_N$, which in our system transforms to 1.6\,$T_{\textrm{BKT}}$ (over the center of the BEC dome), we are left with a narrow temperature range of 1.9--3.5~K, that is, (1.6--3.1)$T_{\textrm{BKT}}$ or (1.2--2.2)$T_N$.

Finally, the 3D prediction of $\xi$ from the $T_{1}^{-1}({\mathbf{q}})$ dependence calculated by QMC for the 32$\times$32$\times$8 size system (Fig.~\ref{Fig_XvsT}) confirms the previous discussion: The increase of $\xi$ due to 3D ordering is barely visible at 1.96 K and shows up clearly at 1.74~K. Although this latter point is very close to $T_N$, the finite size effect limit the simulated $\xi$ value as compared to its expected divergence for an infinite system. On the high temperature side, the 3D $\xi$ continues to track the 2D one very closely at all temperatures, as expected.


\section{Analysis and interpretation of the data}
\subsection{Scaling analysis}

\begin{figure}
	\centering
	\includegraphics[width=\columnwidth]{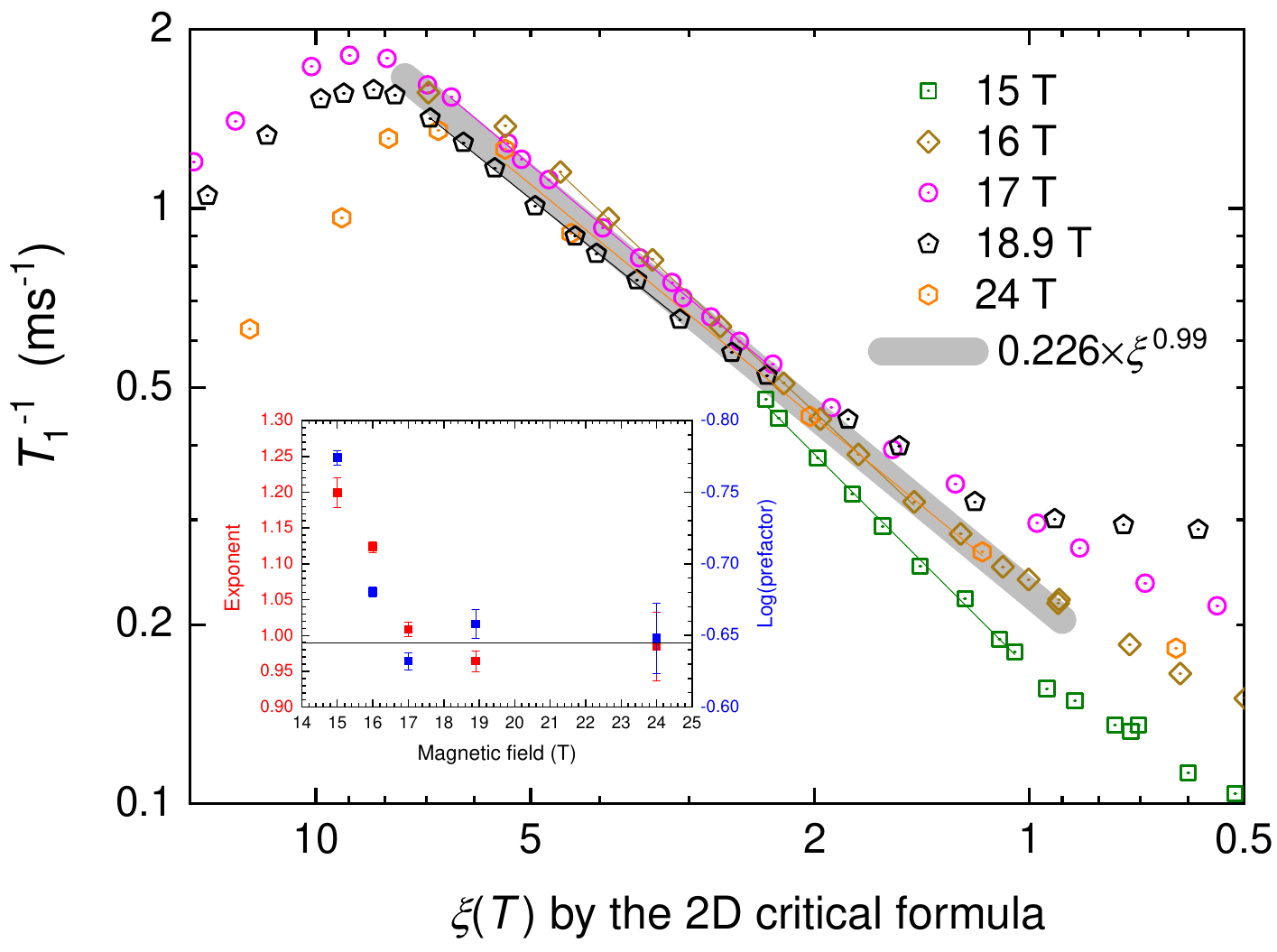}
	\caption{The $T_1^{-1}$ data plotted versus $\xi$ determined by the 2D critical formula $\xi_{\textrm{2D}}(T)$ presents a critical and nearly field-independent behavior in the relevant $\xi$ range. Inset shows the field dependence of the parameters of the corresponding power-law fits given by the thin straight lines. Thick gray line displays the average critical behavior.}
	\label{Fig_CriticalFits}
\end{figure}

In Ref.~\cite{Borsa1992}, Borsa \textit{et al.} provided an interpretation of $T_1^{-1}$ rate that is induced by critical AFM fluctuations, based on scaling arguments. This was derived under the assumption that the fluctuations are enough localized within the Brillouin zone, so the ${\mathbf{q}}$-sum in the Moriya's formula can be replaced by an integral up to infinity over the scaling variable $ x = (\xi a \Delta q)$. For example, in Fig.~\ref{Fig_T1vsQ} we see that this condition is valid at 2.4~K, but certainly not at 7.6 K. When the condition is valid, $T_1^{-1}$ obeys a simple scaling expression
\begin{equation}
\label{eq9}
T_1^{-1}(\xi) = T_0^{-1} \xi^{z-\eta},
\end{equation}
where $ T_0^{-1}$ is the power-law prefactor, $z$ is the classical exponent for the dynamics~\cite{Hohenberg1977,Borsa1992} and $\eta$ the usual BKT critical exponent. To test this scaling on our $T_1^{-1}$ data, we use the $\xi_{\textrm{2D}}(T)$ expression given by Eq.~(\ref{eq8}) to simply convert the $T_1^{-1}(T)$ plot given in Fig.~\ref{Fig_T1vsT} into the $T_1^{-1}(\xi)$ representation, given in Fig.~\ref{Fig_CriticalFits}. Here, we indeed find a power-law scaling behavior within the previously established range of validity of the $\xi_{\textrm{2D}}$ values. Furthermore, we find the data for all the magnetic field values nearly superposed, meaning that $T_0^{-1}$ is practically field independent.

We remark that thus determined scaling exponent is very sensitive to the value of the employed $T_{\textrm{BKT}}$, the relative error in the scaling exponent being 3 times larger than the relative error of $T_{\textrm{BKT}}$. As it is hard to know this latter value to a precision better than 3\%, from our NMR data we conclude that
$z-\eta = 1.0(1)$.

\begin{figure}
	\centering
	\includegraphics[width=\columnwidth]{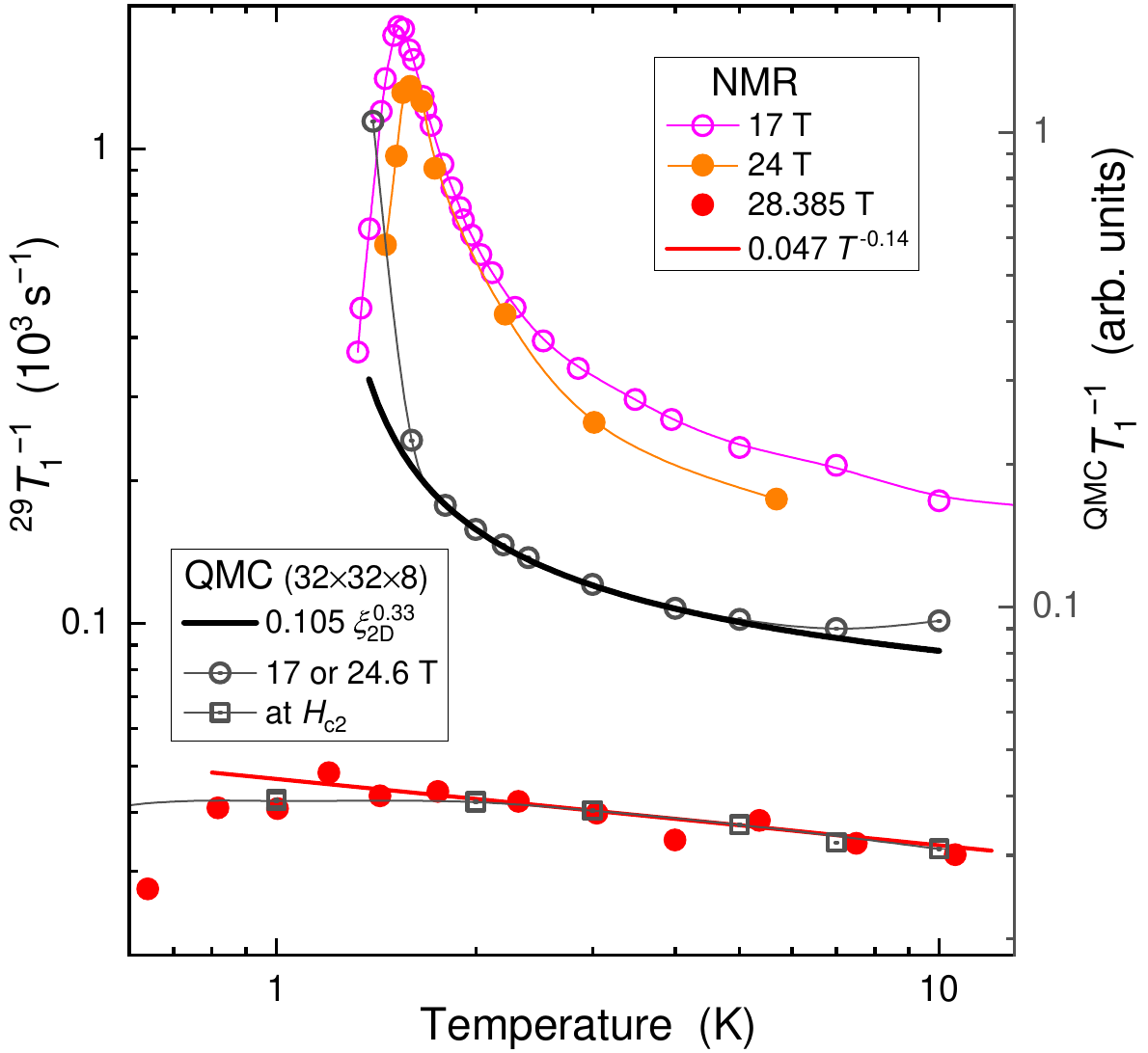}
	\caption{Comparison of the experimental $T_1^{-1}$ data and the corresponding QMC simulations on a 32$\times$32$\times$8 size system. At $H_{c2}$, we find a remarkable overlap of the two quantum critical temperature dependences (red dots and grey squares), $\propto T^{-0.14}$ (red straight line), which allows us to define the relative scale of the two data sets. The QMC data for 17 or 24.6~T (grey circles) then fall much below the corresponding experimental data at 17~T (magenta circles) and 24~T (orange dots) and they display much weaker critical exponent, $T_1^{-1} = 0.105 \xi_{\textrm{2D}}^{0.33}$ (thick black line). Thin lines connecting the data points are a guide to the eye.}
	\label{Fig_T1byQMCandNMR}
\end{figure}

Finally, in Fig.~\ref{Fig_T1byQMCandNMR} we compare the theoretical prediction by QMC of the $T_1^{-1}(T)$ dependence to the NMR data at the critical field and at the field values that are half-way between the critical field and the center of the BEC dome. We first note the perfect overlap of the two temperature dependences at $H_{c2}$, which allows us to establish their relative scaling, which is the relative shift of the two vertical logarithmic scales of this figure. As expected for a quantum critical point, the QMC results {\color{black}shown in the inset in Fig.~\ref{Fig_XvsT}} also confirm that at the critical field the $\xi(T)$ is not of the $\xi_{\textrm{2D}}(T)$ type but is a power law, $\xi(T) \propto T^{-0.655(5)}$, with the exponent very close to $0.67$, the theoretical value of  the 3D XY universality class~\cite{3DXY}.
This scaling relation converts the measured $T_1^{-1}(T) \propto T^{-0.14}$ dependence into $T_1^{-1}(\xi) \propto \xi^{0.21}$. The latter exponent is smaller by 0.8 than the one determined above the BEC dome, clearly indicating the change of the critical regime.

Once the data at $H_{c2}$ allowed us to define the relative scale between the QMC predictions and the NMR $T_1^{-1}$ data, we find that the predicted $T_1^{-1}(T)$ dependence at 17~T (and, equivalently, 24.6~T) falls by more than a factor of 2 below the experimental data, and presents much weaker critical peak. The critical fit to this data, $T_1^{-1} = 0.105 \xi_{\textrm{2D}}^{0.33}$, reveals the exponent 3 times smaller than the one found from the experimental data, pointing to some fundamental mismatch.

Indeed, by performing a careful analysis of our imaginary-time QMC data, we have found that extracting $T_1^{-1}$ may not be very reliable: for instance, we get significantly different results depending whether we use SAC on the local time-dependent spin correlation function or if we perform SAC to extract $T_{1}^{-1}({\mathbf{q}})$ and then sum over $\mathbf{q}$ afterward. This is expected since there is a lot of structure in $\mathbf{q}$ space. We have also tried to use a simpler proxy~\cite{Randeria1992} where $1/T_1$ can be obtained directly from imaginary-time data at the largest time difference [$1/(2T)$], which is often used in the literature. Quite surprisingly, these results are significantly different from the ones obtained using SAC. It would be interesting to investigate this discrepancy in a simpler model and/or using some artificial data for which benchmarks would be available. Last but not least, we are limited to linear size of $L=32$ (with only few points at $L=64$) when computing dynamics, while obtaining equal-time correlation (and then correlation length) can be done on much larger systems (data up to $L=128$ in Appendix A).

\subsection{Critical dynamics}
Let us now discuss the 3D effects. By symmetry, the transition is expected to be in the 3D XY universality class and the spin dynamics has a conserved magnetization along the field, and a nonconserved order parameter in the transverse direction. According to Ref.~\onlinecite{FolkMoser2006}, this corresponds to ``model E'' dynamic, for which the critical NMR properties have been previously discussed~\cite{Dupont2018}, resulting in a diverging $T_{1}^{-1} \sim |T-T_c|^{-0.39}$.

By computing numerically $T_1^{-1}$ both on 2D and 3D lattices, we have checked that there is no significant difference above 1.2$T_N$ typically. Hence, one could only observe 3D criticality in a very small temperature window. However, because of the microscopic distribution of $T_N$ or the cut-off in the mathematical divergence, this regime is inaccessible in our experiments. This is the reason why we only discuss the 2D critical behavior in this manuscript.

Regarding the experimental $z-\eta$~= 1.0(1) value for the 2D critical
exponent, taking the BKT value $\eta = 1/4$, this results in $z=1.25$. We
remark that the classical XY model in 2D  was reported to present the
$z=2$ value for a purely relaxational dynamics, but $z=1.5$ when the
dynamics was modeled by resistively shunted Josephson
junctions~\cite{Jensen2000}. By contrast, for the spin-1/2 quantum XY
model with Hamiltonian dynamics, an exponent $z=2.35$ was
found~\cite{Ying1998}. A reasonable agreement between the experimental value of
$z$ and the theoretical one for resistively shunted Josephson dynamics indicates the
relevance of dissipative interactions between the spins and the lattice.

{\color{black}We also remark that the dependence \mbox{$T_1^{-1} \propto [k_B T/(2 \pi \rho_s)]^{3/2} \xi(T)$}  was derived by Chakravarty and Orbach in Ref. \cite{Chakravarty-Orbach1990} by combining the Moriya formula \cite{Moriya1956} for $T_1^{-1}$ and the dynamical structure factor of a single plane of quantum Heisenberg antiferromagnet \cite{Chakravarty1989}. This system has only a short range order for $T$~$>$~0, with a correlation length $\xi \propto \exp(2 \pi \rho_s / k_B T)$. The given formula, which forces a more restrictive scaling with $z$~=~1 from the outset, is thus not relevant to a finite temperature BKT transition that we address here.}

Finally, from a theoretical point of view, it would be very interesting to characterize the 2D/3D crossover~\cite{Furuya2016} using a simpler microscopic model. However, this remains quite challenging since having different energy scales (inplane vs interplane coupling) would require quite intensive simulations. Below the critical temperature, it would also be interesting to characterize the NMR dynamics in the magnetically ordered phase.

\vspace{7mm}

\section{Conclusions}

This study establishes Ba$_2$CuSi$_2$O$_6$Cl$_2$ as a model compound for exploring BKT criticality in quasi-2D quantum magnets. Combining NMR experiments with QMC simulations, we mapped out the magnetic field-temperature phase diagram and identified the regime dominated by 2D BKT fluctuations. The observed strong and extended peak in the spin-lattice relaxation rate $T_1^{-1}(T)$ dependence confirms the dominance of 2D critical dynamics across a significant temperature window.

However, we observed some discrepancies between our numerical simulations and the experimental NMR relaxation data. While QMC on the effective XXZ model accurately reproduces the BEC phase boundaries from static observables, it fails to capture the full dynamics of the spin system, particularly the magnitude and sharpness of the $T_1^{-1}$ peak in the BKT regime. This might be due to the analytic continuation procedure which requires to extract real-frequency information from imaginary-time QMC data, thus introducing some uncertainty. Finite-size effects may also be another source of limitation, given that computing dynamics is more demanding than focusing on equal-time properties, e.g., correlation length. This clearly demonstrates the complexity of the dimensional crossover from 2D to 3D physics, in particular close to the N\'eel critical temperature $T_N$.

\begin{acknowledgments}
This work has been supported by LNCMI-CNRS, a member of the European Magnetic Field Laboratory (EMFL), by the EUR Grant NanoX No. ANR-17-EURE-0009 in the framework of ''Programme des Investissements d'Avenir'', by the HQI initiative \cite{hqi}, and by France 2030 under the French National Research Agency Award No. ANR-22-PNCQ-0002. We acknowledge the use of HPC resources from CALMIP (grants 2022-P0677 and 2023-P0677) and GENCI (projects A0130500225 and A0150500225). M.D. completed his contribution to this work prior to joining Rigetti Computing.
\end{acknowledgments}

\appendix

\section{QMC computations of ``static'' $\xi$ values}
\label{appA}

Correlation length can be defined using the transverse static structure factor $S_0^{\perp}({\bf q})$ [Eq.~(\ref{eq7})] around its maximal value:
\begin{equation}\label{eq:xi1_qmc}
\xi_a = \frac{1}{q_1}\sqrt{\frac{S_0^{\perp}(\pi,\pi)}{S_0^{\perp}({\bf q_1})}-1},
\end{equation}
where ${\bf q_1}$ is the closest point to $(\pi,\pi)$ at distance $2\pi/L$ in the Brillouin zone. It is possible to get a better approximation~\cite{TodoKato2001} using
\begin{equation}\label{eq:xi2_qmc}
\xi_b = \frac{1}{q_1}\sqrt{\frac{S_0^{\perp}({\bf q_1})-S_0^{\perp}({\bf q_2})}{4S_0^{\perp}({\bf q_2})-S_0^{\perp}({\bf q_1})}},
\end{equation}
where ${\bf q_2}=2 {\bf q_1}$.

\begin{figure}[t]
	\centering
	\includegraphics[width=\columnwidth]{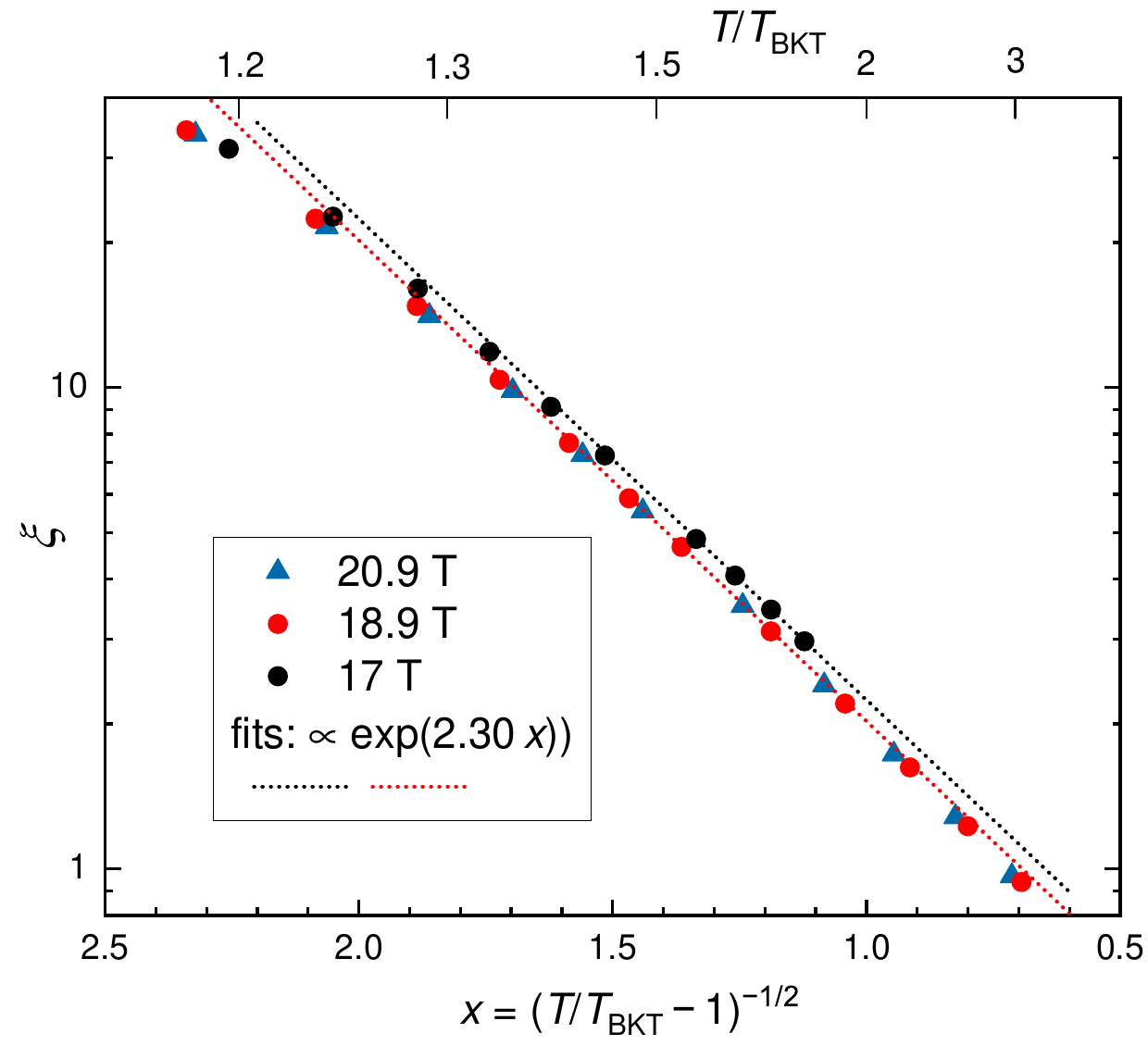}
	\caption{Transverse spin correlation length $\xi_{\textrm{2D}}(T)$ dependence for different magnetic field obtained from QMC simulations, plotted using the $\log(\xi)$ vs. $x = 1/\sqrt{T/T_{\textrm{BKT}}-1}$ scales that are expected to linearize the dependence. For all three magnetic fields, data are very similar and can indeed be fitted using the known scaling function Eq.~(\ref{eq8}).}
	\label{Fig_xi_scaling}
\end{figure}

We have computed these correlations using QMC on 2D square lattices up to linear size $L=128$ and plot some typical data in Fig.~\ref{Fig_xi_scaling}. The temperatures have been normalized using previously determined critical temperatures $T_{\textrm{BKT}}$--see the main text. Quite remarkably, all numerical data collapse onto a single scaling curve which follows the theoretical prediction given by Eq.~(\ref{eq8})~\cite{Kosterlitz1974,Ding1990},
where the numerical value of $b$ found here is 2.30, quite close to the $1.86$ value determined in Ref.~\cite{Laflorencie2012} {\color{black} for a different 2D system. The value of $b$ is not universal but depends on the precise form of the Hamiltonian. For example, in Ref.~\cite{Benfatto2013} it is theoretically predicted that $b$ should depend on the difference between the calculated mean-field transition temperature and $T_{\textrm{BKT}}$. Reported experimental values vary considerably: For example, in chromium triangular lattice antiferromagnets LiCrO$_2$, NaCrO$_2$, and HCrO$_2$, which are frustrated spin systems, the $b$ values determined from the ESR linewidth measurements are respectively 0.52(1), 1.49(2), and 4.1(5) \cite{Hemmida2009}.}

\section{{\color{black}Structural phase transition}}
\label{appB}

\begin{figure}[t]
	\centering
	\includegraphics[width=\columnwidth]{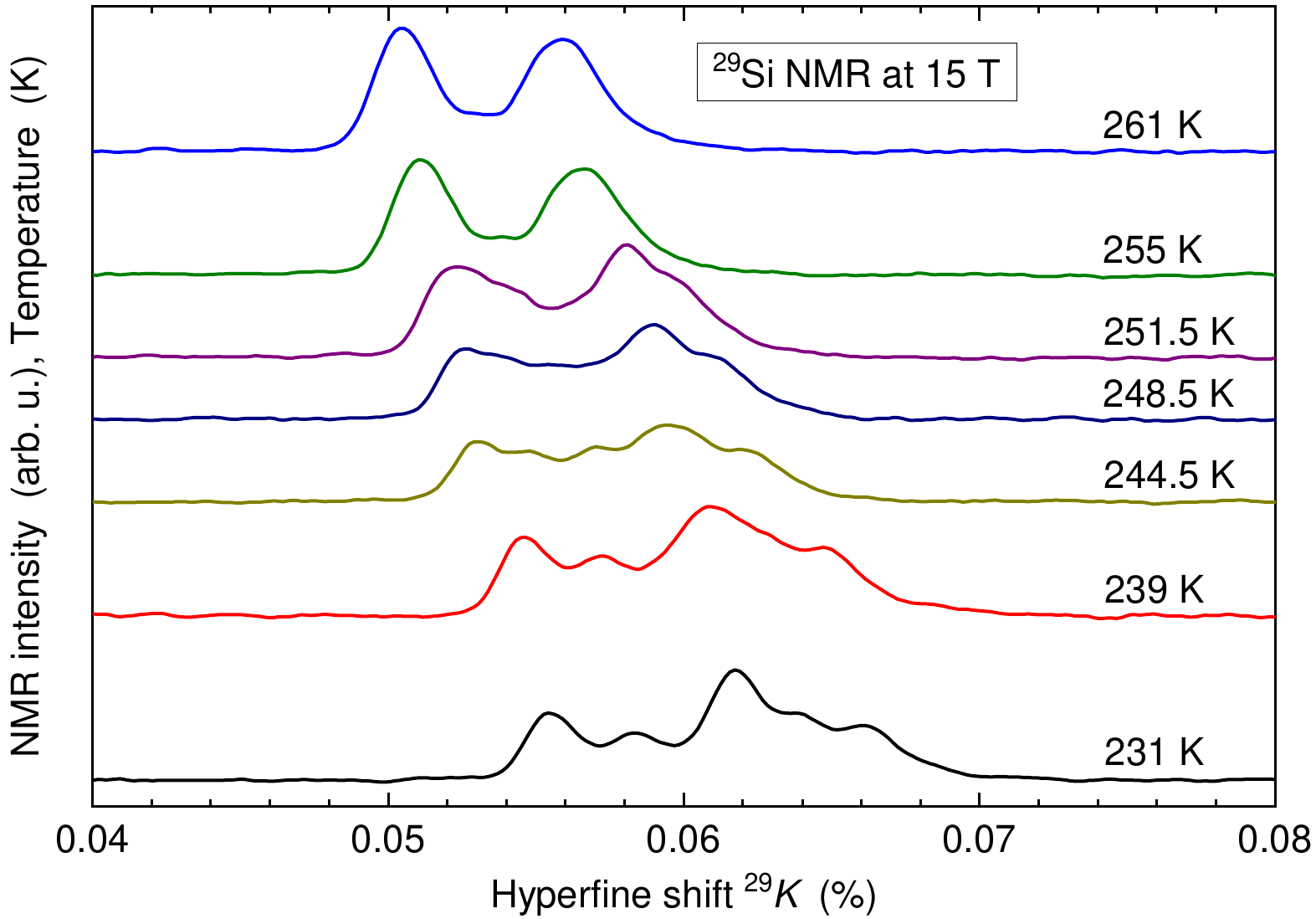}
	\caption{{\color{black}Temperature dependence of $^{29}$Si NMR spectrum reveals a structural phase transition. The spectra are normalized by their integral, and their vertical offset is proportional to the temperature.}}
\label{Fig_PhaseTransition}
\end{figure}

{\color{black}The available structural data have been determined only at room temperature \cite{Okada2016, Nawa2019}, whereas the latter reference finds four different Si crystallographic sites. For the magnetic field applied along the $c$ axis, the $^{29}$Si NMR spectrum should thus present four lines. However, at about 250~K, the temperature dependence of the $^{29}$Si spectrum shown in Fig.~\ref{Fig_PhaseTransition} reveals a splitting of lines that is indicative of a structural phase transition: Only two lines are apparent at 261~K, which probably corresponds to overlapping four lines. On decreasing temperature, in a narrow temperature range at about 250~K appears a line splitting that considerably broadens the spectrum, and at 231~K we clearly observe \emph{more} than four lines. The spectra apparently reflect a doubling of the number of Si sites, leading to clearly separated eight Si NMR lines observed at lower temperatures (inset in Fig.~\ref{Fig_GapedCriticalT1}).}


\section{``Dynamic'' and ``static'' correlation lengths}
\label{appC}

We consider the ansatz that is compatible with Eqs. (\ref{eq5}--\ref{eq7})
\begin{equation}\label{eq:S0}
S_{\mathbf{q}}^{\perp}(\omega=0) \propto \xi^{z+2-\eta} \exp[-\sqrt{(\xi \Delta{\mathbf{q}})^2 + d^2} + d],
\end{equation}
which defines the ``dynamic'' $\xi$ (here in units of $a$) as determined by QMC from the $T_1^{-1}(\mathbf{q})$ simulations; see Fig.~\ref{Fig_T1vsQ}.
Into this expression we introduce an \textit{ad hoc} frequency dependence that satisfies the required scaling
\begin{equation}\label{eq:S0omega}
S_{\mathbf{q}}^{\perp}(\omega) \propto \xi^{z+2-\eta} \exp[-\sqrt{(\xi \Delta{\mathbf{q}})^2+ (|\omega|^{1/z} \xi \lambda)^2 + d^2} + d].
\end{equation}
Integrating over the frequency we then obtain the $t = 0$ (equal-time) correlator, i.e., the static structure factor, used to determine by QMC the ``static'' $\xi$ values
\begin{equation}\label{eq:S0integral}
S_0^{\perp}(\mathbf{q}) \propto \int_{-\infty}^{+\infty}S_{\mathbf{q}}^{\perp}(\omega) d\omega.
\end{equation}
For the expression given by the Eq.~(\ref{eq:S0omega}), this integral can be evaluated analytically in terms of the modified Bessel function
of the second kind $K_{\nu}(x)$, leading to the following $\mathbf{q}$ dependence
\begin{equation}\label{eq:S0q}
S_0^{\perp}(\mathbf{q}) \propto \sqrt{(\xi \Delta{\mathbf{q}})^2 + d^2}^{\frac{z+1}{2}}K_{\frac{z+1}{2}}(\sqrt{(\xi \Delta{\mathbf{q}})^2 + d^2}).
\end{equation}
A power series expansion
\begin{equation}\label{eq:S0series}
S_0^{\perp}(\mathbf{q})/S_0^{\perp}(0) = 1 - (r_{\xi} \xi \Delta{\mathbf{q}})^2 + \textrm{O}(\Delta{\mathbf{q}})^4
\end{equation}
defines the small $\Delta{\mathbf{q}}$ behavior and thus the corresponding static correlation length value $r_{\xi} \xi$, where
\begin{equation}
\label{eq:ratio}
r_{\xi} = \sqrt{[z+1-d K_{\frac{z+3}{2}}(d)/K_{\frac{z+1}{2}}(d)]/(2 d^2)}
\end{equation}
is the ratio between the dynamic and static $\xi$ values. For the experimentally determined values of $z=1.25$ and $d=0.82$, it evaluates to $r_{\xi}=0.60$, very close to the value 0.575 shown in Fig.~\ref{Fig_XvsT}, obtained from the comparison of the two values as determined from the two types of QMC simulations.


\end{document}